\documentclass[runningheads]{llncs}
\usepackage{afterpage}
\usepackage{algorithm,algorithmic}
\usepackage{array} 
\usepackage{amsmath,amssymb,amsfonts}
\usepackage{booktabs} 
\usepackage{color}
\usepackage{graphbox}
\usepackage{graphicx}

\usepackage{mathtools}
\usepackage{multirow}
\usepackage{tabularx}
\usepackage{textcomp}
\usepackage{tikz}
\usepackage{pifont} 
\usepackage{colortbl}
\usepackage[normalem]{ulem}
\useunder{\uline}{\ul}{}
\usepackage{hyperref}
\usepackage{cleveref} 
\usepackage{atbegshi} 

\usepackage{bibentry}
\usepackage{longtable}
\usepackage{csquotes}

\usepackage{flushend}
\usepackage{cleveref}
\usepackage{subfig}
\usepackage{graphicx}
\usepackage{orcidlink}
\usepackage{changepage}

\usepackage{pgfplots}
\usepackage{src/tkz-kiviat} 
\usetikzlibrary{arrows}

\usepackage[nomessages]{fp}
\usepackage{censor}
\usepackage[font=small,skip=6pt]{caption}
\usepackage{lmodern}
\usepackage{relsize}

\newcounter{ch}
\usepackage{array,ragged2e}
\newcolumntype{P}[1]{>ŵ{\RaggedRight\arraybackslash}p{#1}}
\usepackage[T1]{fontenc}
\AtBeginShipout{\ifnum\value{page}=16\pagecolor{red!20}\fi}
\begin{document}
\title{SLO-Aware Task Offloading within\\Collaborative Vehicle Platoons\thanks{Partially funded by the National Natural Science Foundation of China (NSFC) under grant numbers 52220105001 and 72322002, and partially by the European Union under (TEADAL, 101070186) and (INTEND, 101135576)}}
\titlerunning{SLO-Aware Task Offloading within Collaborative Vehicle Platoons}
%


\author{Boris Sedlak\inst{1}\orcidlink{0009-0001-2365-8265} \and
Andrea Morichetta\inst{1}\orcidlink{0000-0003-3765-3067}
\and
Yuhao Wang\inst{2}\orcidlink{0000-0001-8283-8971}
\and
Yang Fei\inst{2}\orcidlink{0000-0002-4342-2196}
\and
Liang Wang\inst{2}
\and
Schahram Dustdar\inst{1}\orcidlink{0000-0001-6872-8821} \and
Xiaobo Qu\inst{2}\orcidlink{0000-0003-0973-3756}}
\authorrunning{Sedlak et al.}
%
\institute{Distributed Systems Group, TU Wien, Vienna, Austria. \and
Emerging Transportation Solutions, Tsinghua University, Beijing, China
}

\maketitle              
\begin{abstract}

In the context of autonomous vehicles (AVs), offloading is essential for guaranteeing the execution of perception tasks, e.g., mobile mapping or object detection. While existing work focused extensively on minimizing inter-vehicle networking latency through offloading, other objectives become relevant in the case of vehicle platoons, e.g., energy efficiency or data quality for heavy-duty or public transport. Therefore, we aim to enforce these Service Level Objectives (SLOs) through intelligent task offloading within AV platoons. We present a collaborative framework for handling and offloading services in a purely Vehicle-to-Vehicle approach (V2V) based on Bayesian Networks (BNs). Each service aggregates local observations into a platoon-wide understanding of how to ensure SLOs for heterogeneous vehicle types. With the resulting models, services can proactively decide to offload if this promises to improve global SLO fulfillment. We evaluate the approach in a real-case setting, where vehicles in a platoon continuously (i.e., every 500 ms) interpret the SLOs of three actual perception services. Our probabilistic, predictive method shows promising results in handling large AV platoons; within seconds, it detects and resolves SLO violations through offloading.



\end{abstract}


\section{Introduction}\label{sec:introduction}


The swift evolution of Autonomous Vehicles (AVs) promises a disruptive impact~\cite{ma2020artificial} for future transportation. From car sharing to heavy-duty or public transport, AV solutions claim considerable benefits, such as rapid green transition and traffic flow improvement~\cite{kuutti2020survey}. At the same time, the execution of AV-enabling services, such as perception, path
planning, and control~\cite{le2022survey} pose ambitious processing requirements. Here, optimal allocation and execution of workloads highly depend on AVs' constrained computation capabilities and 
the supporting infrastructure's network bandwidth.
For perception services, a lack of computation resources can cause delays in real-time decision-making, leading to potentially harmful consequences. Hence, offloading services (or tasks) to other computing entities releases the pressure on AVs' resources.

Typically~\cite{fan2023joint}, service offloading minimizes latency between neighboring vehicles through Vehicle-to-Vehicle (V2V) or Vehicle-to-Infrastructure (V2I) communication.
However, collaborative AV scenarios commonly have higher-level objectives besides latency.
For instance, consider AV platoons for public or heavy-duty transport, where the system providers might want to optimize vehicle paths, save energy, or minimize cost.
We define these requirements as Service Level Objectives (SLOs), borrowing the term from software engineering. The concept of SLOs is wide enough to define any high- or low-level objective that a management framework can enforce~\cite{sedlak_equilibrium_2024} by elastically adapting hardware or software. SLO-awareness offers promising scenarios~\cite{qiao2018collaborative} for V2V offloading; still, its adoption remains limited. This gap highlights the potential for developing more intelligent offloading mechanisms~\cite{guo_toward_2020}.


This work, therefore, aims to ensure SLOs by incorporating them into the offloading mechanism -- we call this ``SLO-aware task offloading''. Our motivation stems from two central objectives: (1) we want to ensure that vehicles fulfill the SLOs of their local services; if SLOs are violated, this might be resolved by offloading services, and simultaneously, (2) offloaded tasks must not jeopardize the SLO fulfillment of existing services at the target host.
This goal implies solving a combinatorial problem, i.e., the optimal assignment of $n$ services to $m$ vehicles. However, this problem is NP-hard, hence practically intractable. A solution could be to decompose the problem by bounding the service offloading decision within single AVs. Even so, this approach is ineffective as training an offloading model for every AV separately would introduce a considerable overhead. Furthermore, we would miss the chance to combine the knowledge from multiple AVs, which promises a more profound understanding due to larger numbers of observations. For these reasons, we envision a method that trains a decision model within an AV but simultaneously integrates knowledge from other AVs.

In this paper, we present a modular, collaborative framework for autonomous SLO interpretation and service offloading. Here, we consider collaborative offloading approaches using ``decentralized'' sensory data~\cite{gao2021federated}.
Individual services continuously observe their processing to understand the extent to which SLOs can be fulfilled on different processing hardware; this knowledge is encoded in an SLO interpretation (SLO-I) model. 
These models are updated by a mutable platoon leader according to AVs' observations and then broadcast to other AVs. Given the SLO-I model, individual services predict how offloading would impact global SLO fulfillment. 
Hence, the contributions of this article are: 
\begin{enumerate}
    \item An SLO-aware offloading mechanism based on Bayesian networks that dynamically estimates the hardware implications of multiple competing services to find a satisfying assignment. Thus, it is possible to optimize the SLO fulfillment by shifting computation within a composable vehicle platoon.
    
    \item A collaborative training strategy that continuously exchanges model updates between edge devices while adjusting the training frequency according to agents' local SLO prediction errors. Thus, service agents improve their SLO interpretation whenever the system does not behave as predicted.
    
    \item A modular framework for collaborative service offloading that can be extended with custom processing services and respective SLOs. Thus, other service managers can plug their own service implementation into the framework, which itself can be installed on arbitrary edge device types.
\end{enumerate}

The remainder of this paper is organized as follows: Section~\ref{sec:relatedwork} further illustrates the scenario used throughout this paper and gives an overview of related work, Section~\ref{sec:methodology} describes our framework for SLO-aware offloading, which is evaluated in Section~\ref{sec:evaluation}. Finally, we summarize our paper in Section~\ref{sec:conclusion}.

\section{Preliminaries}\label{sec:relatedwork}
To highlight the need for our methodology and the research gap it fills, this section first presents an illustrative scenario in which SLO awareness improves high-level requirements fulfillment within a vehicle platoon. Further, we discuss to what extent these problems have been addressed in existing research.

\subsection{Illustrative Scenario}
\label{subsec:illustrative-scenario}




\begin{figure}[t]
\centering
\subfloat[Maneuvers of the platoon]{\label{subfig:platoon-maneuver}\centering\includegraphics[width=.45\columnwidth]{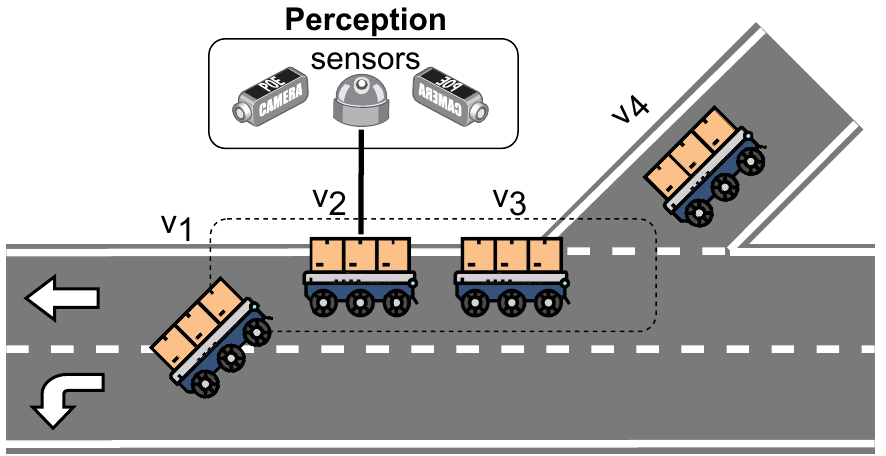}}
\subfloat[Offloading within the platoon]{\label{subfig:platoon-placement}\centering\includegraphics[width=.45\columnwidth]{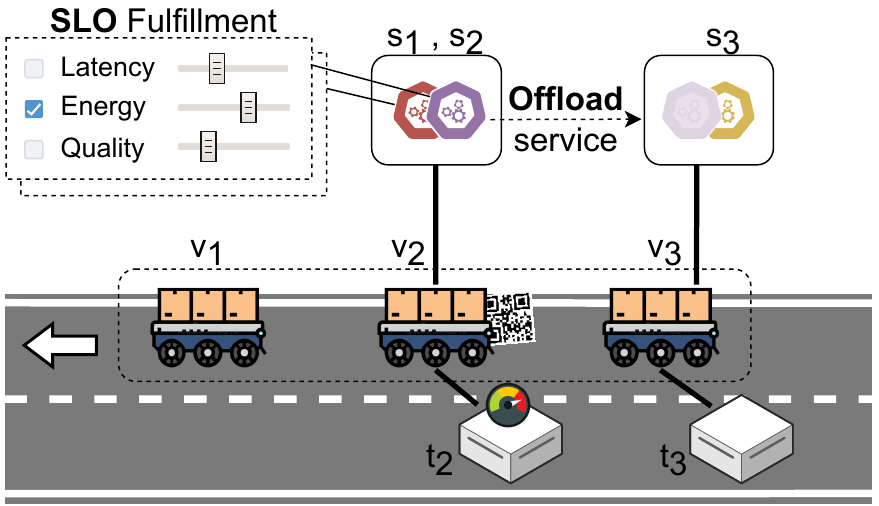}
\label{fig:rebalance}}
\hfill
\caption{Composite vehicle platoons offload computations according to SLO fulfillment; if service $s_2$'s SLOs are not fulfilled at host $v_2$, it searches for alternatives, such as $v_3$}
\label{fig:platoon-scenario}
\end{figure}

Here, we consider a platoon of vehicles for heavy-duty transportation. Depending on the trajectories of platoon members, individual vehicles can join or leave the platoon at specific intersections, such as ramps.
In our work, we focus on V2V offloading, as V2I infrastructures could be impractical~\cite{fan2023joint} or add delays~\cite{chen2020delay}.

In detail, as shown in Fig.~\ref{subfig:platoon-maneuver}, $n$ vehicles are clustered into a platoon $P=\{v_1,...,v_n\}$. We represent each vehicle through the pair $v = \langle id, t \rangle$, where $v.t$ specifies the type of processing device embedded. 
Plus, each vehicle is equipped with numerous sensor systems and perception services. For instance, in Fig.~\ref{subfig:platoon-placement}, vehicle $v_2$ runs two services, i.e., mapping its surroundings through Lidar ($s_1$) and detecting objects on the road through computer vision ($s_2$). Given that $v_2$ has a QR code attached to its rear, $v_3$ follows its predecessor by scanning for QR codes ($s_3$). We define a service through $s = \langle type, Q, C \rangle$, which reflects the $type$ of perception service, e.g., Lidar or CV; $Q$ specifies a set of processing SLOs, and $C$ a list of service constraints, e.g., CV should operate at $\textit{fps} = 15$. 
Given this setup, the operator specifies clear SLOs on how fast the vehicles must respond to dynamic road conditions, which improves safety during operation.

Depending on services' resource demand, vehicles may not possess sufficient processing capabilities to fulfill their SLOs, which impacts the latency and quality of how a vehicle perceives its environment. For instance, $v_2$ might employ a weaker processing device ($v_2.t$); however, $v_3$'s resources are less utilized, so $v_2$ might offload one of its services to $v_3$.
Therefore, $v_1$ must now decide (1) which service, i.e., $s_1$ or $s_2$, should best be offloaded to $v_3$, (2) whether this improves SLO fulfillment of remaining services at $v_2$, and (3) if offloading could impact $s_3$ negatively. In the context of this paper, 
we focus on higher-level requirements, i.e., leaving out networking latencies for transferring input data and results under the assumption of high network throughput between nearby vehicles.

\subsection{Related Work}

We classify existing literature on task offloading for IoV and related scenarios in two main categories: offloading in V2I / V2V scenarios and offloading through Markovian or Bayesian methods. To set the foundation for our contribution, we highlight the strengths and limitations of these approaches.

\vspace{-10pt}
\subsubsection{IoV offloading mechanisms}
In the context of V2I task offloading, Xu et al.~\cite{xu_optimization_2023} provide a neighborhood search algorithm that minimizes costs of task outsourcing, estimated on simulated network traffic. 
Similarly, Dong et al.~\cite{dong_quantum_2023} provide a multi-task and multi-user offloading mechanism for Mobile Edge Computing (MEC), optimized through a particle swarm. 
Ant colony optimization (ACO) is another explorative algorithm for optimal pathfinding:
Mousa and Hussein~\cite{mousa_efficient_2022} apply ACO to cluster IoT devices accessed by UAVs; 
Ma et al. \cite{ma_edge_2022} model the same scenario, but with Mixed-Integer Linear Programming (MILP), closely to Zhang et al.~\cite{zhang_distributed_2023}.
Related to our use case, Lu et al.\cite{lu_cooperative_2022} provide a latency-aware V2V/V2I offloading mechanism based on Deep Reinforcement Learning (DRL). 
Fan et al.~\cite{fan2023joint} propose a V2V/V2I offloading tool that decomposes optimization problems with Generalized Benders Decomposition (GBD).

Other authors model offloading scenarios as shortest path~\cite{fan_minimum-cost_2019} or stochastic optimization problem~\cite{hu_task_2019};
some methodologies focus on \textbf{solely V2V} offloading: Du et al.~\cite{du_cooperative_2020} provide a collaborative offloading mechanism for sensing tasks in autonomous vehicle platoons, making use of idle resources. 
Guo et al.~\cite{guo2022v2v} combine LSTM-based trajectory prediction and optimization strategy for V2V offloading. 
However, all these methods, while solid, rely on simulations rather than real-world data, assume static and homogeneous infrastructures, which are unrealistic, and frequently neglect SLO measures like energy consumption.

\vspace{-10pt}
\subsubsection{Offloading through Markovian and Bayesian methods}
To the best of our knowledge, there are no solutions based on Bayesian Networks for V2V task offloading in platoons. Still, Markov models and Bayesian approaches are found in Edge-to-Cloud scenarios for task offloading~\cite{sedlak_equilibrium_2024,sedlak_diffusing_2024}.
Hazra et al.\cite{hazra_cooperative_2023} use MILP to find offloading locations in hierarchical computing environments under latency and energy constraints. 
Wu et al.\cite{wu_intelligent_2023} offload streaming tasks from edge nodes to fog or cloud resources through a Markov decision process, improved through Reinforcement Learning (RL). 
Tasoulas et al. \cite{tasoulas_bayllocator_2012} provide a prediction mechanism that uses historical observations to forecast VMs' resource demand through Bayesian Networks. 
However, these papers offer little variety for SLOs and do not incorporate dynamic or real-time adaptations.

\vspace{-10pt}
\subsubsection{Takeaways}
Existing research focused extensively on MEC offloading mechanisms to RSUs or UAVs for optimizing network latency. Despite being an essential metric in vehicle control, AVs can have more specific objectives, e.g., energy efficiency for public or heavy-duty transportation, or quality in case of mobile mapping.
%
%
In addition, most approaches were only evaluated theoretically; however, to establish reliable offloading mechanisms, it is paramount to consider dynamic runtime behavior. 
Conversely, we propose an SLO-aware mechanism for V2V offloading that optimizes various types of SLOs in heterogeneous vehicle platoons. Centralized approaches suffer from the combinatorial complexity of finding a globally optimal solution and the risk of becoming a single point of failure; in our approach, however, services themselves get decentralized authority to interpret their runtime behavior and make offloading decisions.


\section{Methodology}\label{sec:methodology}
\begin{figure}[t]
\centerline{\includegraphics[width=0.9\columnwidth]{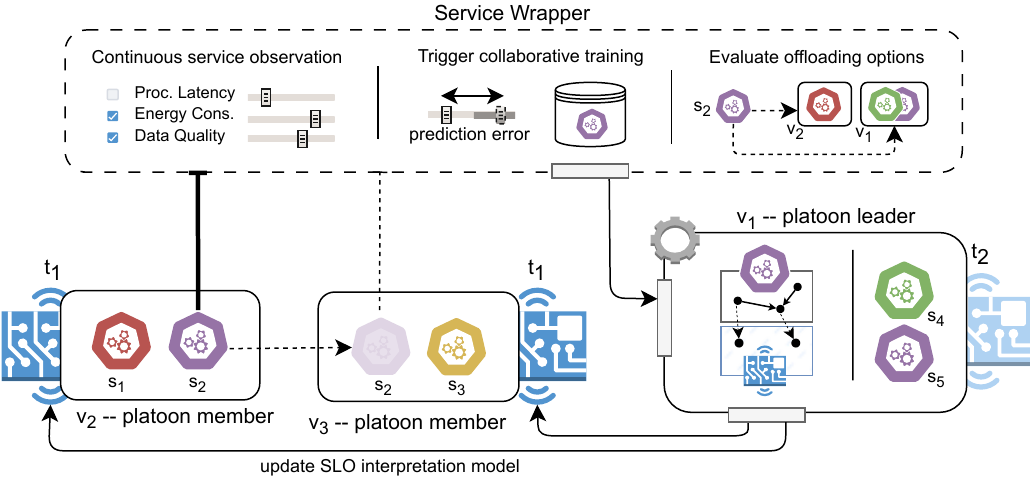}}
\caption{Framework for collaborative offloading: inaccurate SLO predictions trigger retraining of SLO interpretation models; services use these models to evaluate alternative hosts according to their expected hardware utilization and SLO fulfillment}
\label{fig:high-level-methodology}
\end{figure}

In the following, we present our modular framework for SLO-aware task offloading in composable vehicle platoons. This means, continuously observing service executions to collect insights, interpreting these insights through collaborative training, and making offloading decisions. Fig.~\ref{fig:high-level-methodology} provides a high-level overview of these processes, which are explained in more detail in subsections \ref{subsec:service-observation} to \ref{subsec:service-offloading}.

\subsection{Service Observation}
\label{subsec:service-observation}

The first building block of our approach is observing a service, i.e., continuously monitoring and interpreting its SLO fulfillment.
Observation requires interpreting service metrics parallel to service execution, as part of the service wrapper in Fig.~\ref{fig:high-level-methodology}. Perception tasks, such as those executed by autonomous vehicles, usually work iteratively; hence, service metrics are also interpreted step by step. In Algo.~\ref{alg:observation}, it is depicted how metrics ($D_{s,v}$) from executing a service ($s$) on a vehicle ($v$) are interpreted: for a set of SLOs ($Q$), the percentage of metrics ($\phi$)\footnotemark that fulfill these conditions is determined as shown in Eq.~\eqref{eq:slo-f}; then, $\phi$ is appended to the sliding window $W_\phi$. To avoid overhasty decisions based on sporadic SLO violations, the length of the sliding window ($|W_\phi|$) can be customized.

\begin{subequations}
\label{eq:slo-f}
    \begin{equation}
        \label{eq:Q-phi}
        \phi(Q) = \frac{\sum_{i = 1}^{|Q|} \phi(q_{i})}{|Q|}
    \end{equation}    
    \begin{equation} 
        \label{eq:q-phi}
        \phi(q_{i}) = \phi(q_{i}, m, v|\forall m \in {D_{q_{i}}^{v}}, v \in {V}) =
            \sum_{j=1}^{|D_{q_{i}}^{v}|}{
                \frac{\phi(m_{j}, q_{i})}{|D^{v}|}
            }
    \end{equation}
    \begin{equation}
        \label{eq:q-phi-m}
        \text{where }\phi(q_{i}, m_{j}) = 
        \begin{cases}
        1,& \text{if}\ m^{q_{i}}_{j_{\text{min}}} \leq m_{j} \leq m^{q_{i}}_{j_{\text{max}}} \\
        0,& \text{otherwise}
        \end{cases}
    \end{equation}
\end{subequations}
\vspace{1pt}

\footnotetext{We choose the symbol $\phi$ due to the sound of the letter, i.e., SLO ful-phi-llment}

To understand if a service should be loaded off, we consider both its current SLO fulfillment as well as predictions according to historical observations; for this, we infer the predicted SLO fulfillment (Line 3) using a Bayesian Network (BN).
BNs are structural causal models encoded as Directed Acyclic Graph (DAG); nodes represent random variables (e.g., $cpu$) and an edge between two variables (e.g., $cpu \rightarrow energy$) indicates conditional dependency, i.e., $cpu$ influences the states of $energy$. 
%
Given historical observations, BNs can answer how likely it is to observe a specific (i.e., SLO fulfilling) state at runtime~\cite{sedlak_equilibrium_2024,sedlak_diffusing_2024};
%
hence, we call them SLO interpretation (SLO-I) models. For an SLO-I model $m$ and service $s$, agents predict SLO fulfillment through $\texttt{INFER}(m, s.Q, s.C)$.
\begin{equation}
\label{eq:buffer-full}
    \texttt{FULL}(B) = \frac{\sum_{i=1}^{n} 1}{|B|}
\end{equation}
To ensure that predictions remain accurate regardless of variable drifts, increasing prediction errors trigger retraining. As more training data is collected (Line 4), the utilization of the metrics buffer, as shown in Eq.~\eqref{eq:buffer-full}, indicates that the model becomes outdated, putting additional weight on retraining.
Thus, in Line 5, we calculate the evidence to retrain ($e_r$) as the sum of absolute prediction error and metric buffer utilization. If $e_r$ surpasses the retraining rate ($\rho$), the metrics buffer is sent to the platoon leader to update the SLO-I model; this is further elaborated in Section~\ref{subsec:collaborative-training}. Notice that both the maximum buffer size ($|B|$) as well as $\rho$ can be customized; for instance, $\rho = 1.0$ would be exceeded if $\texttt{FULL}(B) = 0.8$ and the prediction is off by $0.3$.

\begin{algorithm}[t]
\caption{Continuous SLO Interpretation}\label{alg:observation}
\begin{algorithmic}[1]
\REQUIRE $D$, $B$, $W_\phi$; $s$, $m_{s,t}$, $\rho$, $\omega$, $\gamma$ (global) 
\STATE $\phi_s \gets \phi(s.Q)$
\STATE $W_\phi \gets W_\phi \cup \phi$
\STATE $p_\phi \gets \texttt{INFER}(m_{s,t}, s.Q, s.C)$
\STATE $B \gets B \cup D$
\STATE $e_r \gets \textbf{abs} (W_\phi - p_\phi) + \texttt{FULL}(B)$
\IF{$e_r > \rho$}
\STATE $m_{s,t} \gets \texttt{RETRAIN}(B); B \gets \emptyset$ 
\ENDIF
\STATE $e_o \gets \textbf{abs} (W_\phi - p_\phi)) + (1 - W_\phi)$
\IF{$e_o > \omega$}
\STATE $v' \gets \texttt{FIND\_OFFLOAD}(s, v)$ 
\STATE \algorithmicif\ $v' \neq \emptyset$ \algorithmicthen\ $\texttt{OFFLOAD}(s,v')$
\ENDIF
\end{algorithmic}
\end{algorithm}

Model retraining assures that offloading decisions are always taken based on accurate assumptions; to that extent, the evidence to load off ($e_o$) is calculated similarly (Line 9) by computing the absolute prediction error and adding the absolute SLO violation. When $e_o$ surpasses a custom rate $\omega$, and only in this case, does the agent look for a suitable host within the vehicle platoon (Line 11); given that there is one, the service will then be offloaded there; this will be explained in more detail in Section~\ref{subsec:service-offloading}. 

\subsection{Collaborative Training}
\label{subsec:collaborative-training}

Retraining of SLO-I models is carried out by the platoon leader, i.e., a distinguished platoon member who was elected for this role; however, the necessary training data can be provided by any platoon member. For instance, recall Fig.~\ref{fig:high-level-methodology}, where $s_2$ and $s_5$ are two instances of the same service, e.g., a CV task, though executed on different hosts. Each service collects evidence to retrain ($e_r$) independently of other instances; once a service collects enough evidence, it sends a model update request to the platoon leader, including its local training buffer. Technically, our architecture allows platoon members to update SLO-I models locally; however, limiting the training to the leader improves the model update consistency over the platoon, plus it helps isolate the training overhead. 

Each combination of service and device type is encoded in a unique SLO-I model. Therefore, as soon as the platoon leader ($v_1$) receives a metric buffer ($B_{s,v_2}$) from a member ($v_2$), it first checks $v_2$'s type of processing device ($v_2.t$), e.g., Jetson Orin NX. Next, the leader locally updates the SLO-I model ($m_{s,t}$) for service $s$ and device type $v_2.t$; in our example, this means updating the SLO-I model of service $s=\textit{CV}$ executed on device type $t=\textit{NX}$. 
%
Finally, a new model version $m' = \texttt{PARL}(m, B)$ is created by updating the BN parameters according to recent observations ($B_{s,v}$).
Retraining through \texttt{PARL} is limited to updating the conditional probabilities of BN variables; the structure (i.e., variable relations) is left untouched and only supplied through expert knowledge.

After retraining, the updated model ($m'$) is shared within the platoon. For this, the platoon leader broadcasts $m'_{s,t}$ to all members in $\{v \in P \mid v.t = v_2.t\}$, i.e., to all the platoon members with the matching device type. Any vehicle that thus receives an updated model now replaces the SLO-I models of its locally running services. For Fig.~\ref{fig:high-level-methodology}, this means that $s_2$ gets updated, but $s_5$ is not, since $v_1$ uses a different processing device (i.e., $v_2.t \neq v_1.t$). Thus, all instances of the CV service now interpret their SLO fulfillment according to the new model version.
While it is possible to limit updates only to devices that currently execute this service type, we chose this variant because it allows vehicles to start executing a service without first having to pull the latest model version.


\subsection{Service Offloading}
\label{subsec:service-offloading}

\begin{algorithm}[t]
\caption{Evaluating Alternative Host (\texttt{FIND\_OFFLOAD})}\label{alg:offloading}
\begin{algorithmic}[1]

\REQUIRE $s$, $v$; $P$, $A$, $M$ (global)
\ENSURE $v'$ \COMMENT{Optimal vehicle for offloading $s$ from $v$}
\STATE \algorithmicif\ $|P| = 1$
\algorithmicthen\ \textbf{return} $\emptyset$

\STATE $S_v \gets \{ s_a \mid (s_a,v_a) \in A \mid  v_a = v\}$
\STATE $S_{v}' \gets S_v \setminus \{s\}$; $\Gamma \gets \emptyset$

\STATE $\phi_{S} \gets \texttt{INFER}(M[S_v], S_v.Q, \texttt{CONV\_HW}(S_v, v.t))$
\STATE $\phi_{S'} \gets \texttt{INFER}(M[S_{v}'], S_{v}'.Q, \texttt{CONV\_HW}(S_{v}', v.t))$

\FOR{\textbf{each} $w$ \textbf{in} $P \setminus \{\textit{v} \}$}
\STATE $\Sigma_w \gets \{ s_a \mid (s_a,v_a) \in A \mid  v_a = w\}$
\STATE $\Sigma_{w}' \gets \Sigma_w \cup \{s\}$

\STATE $\phi_{\Sigma} \gets \texttt{INFER}(M[\Sigma_w], \Sigma_w.Q, \texttt{CONV\_HW}(\Sigma_w, w.t))$
\STATE $\phi_{\Sigma'} \gets \texttt{INFER}(M[\Sigma_{w}'], \Sigma_{w}'.Q, \texttt{CONV\_HW}(\Sigma_{w}', w.t))$
\STATE $\gamma \gets (\phi_{S'} + \phi_{\Sigma'}) - (\phi_S + \phi_\Sigma)$
\STATE $\Gamma \gets \Gamma \cup (\gamma, w)$
\ENDFOR
\STATE $\gamma, v' \gets \{(\gamma, w) \in \Gamma, \textbf{max}(\gamma)\}$
\RETURN $v' \textbf{ if } \gamma > 0 \textbf{ else } \emptyset$
\end{algorithmic}
\end{algorithm}

Given that a service collected sufficient evidence to load off ($e_o$), like $s_2$ in Figs.~\ref{fig:platoon-scenario}~\&~\ref{fig:high-level-methodology}, the service looks for the best alternative assignment, which means comparing for each of the other platoon members if global SLO fulfillment would be improved through offloading there.
Formally, this is described in Algo.~\ref{alg:offloading}, which uses the list of platoon members ($P$), the assignments ($A$) of which vehicle currently executes which service, and the shared collection ($M$) of all SLO-I models. In case the platoon does not contain other vehicles (Line 1), the search stops immediately; otherwise, the service predicts (a) the combined SLO fulfillment ($\phi_S$) for all services ($S_v$) currently executed locally at vehicle $v$ (Line 4), and (b) how offloading $s$ would change local SLO fulfillment ($\phi_{S’}$) (Line 5). For this, we first estimate the combined hardware demand (\texttt{CONV\_HW}) that would emerge from co-locating the services on a target device and then estimate per service if the increased hardware load has an impact on its SLO fulfillment.

Before continuing Algo.~\ref{alg:offloading}, we briefly explain $\texttt{CONV\_HW}(S,t)$, which predicts the hardware utilization that would result from executing all $s \in S$ at a device of type $t$.
For each service $s \in S$, we use the respective model $m_{s,t} \in M$ to infer its expected hardware utilization; in our case, we consider the hardware variables $hw = \{\textit{cpu}, \textit{gpu}, \textit{memory}\}$, but the list can be extended arbitrarily with other monitor variables included in the SLO-I model. This returns a probability distribution (e.g., $p_{cpu}$) for each variable $\in hw$; afterward, the combined hardware load is calculated as the convolution of the individual loads. Formally, the convolution of two or more random variables $(X,Y)$ with probability density functions $f_X(x)$ and $f_Y(y)$, i.e., the probabilities for each \textit{hw} variable, is the sum ($Z = X + Y$) of their individual distributions~\cite{bacchus_representing_1989}, as shown in Eq.~\eqref{eq:convolution}.
\begin{equation}
    \label{eq:convolution}
    f_Z(z) = (f_X * f_Y)(z) = \int_{-\infty}^{\infty} f_X(t) f_Y(z - t) \, dt
\end{equation}
Thus, we obtain the combined hardware utilization, which is supplied as a constraint to $\texttt{INFER}$; this allows estimating how the respective hardware load would impact SLO fulfillment ($\phi_S$ and $\phi_{S'}$). Alternative approaches to estimating combined load and resulting SLO fulfillment might need to empirically test the service deployment, which is infeasible when decisions must be made quickly.

In the next step, we estimate for each of the other platoon members ($w$) the SLO fulfillment ($\phi_\Sigma$) of its local services ($\Sigma_w$) and how this would be affected ($\phi_{\Sigma'}$) if we would offload $s$ there. This follows the same pattern as described for the source vehicle $v$: we use the list of services currently executed at $w$ (Line 7) and their respective SLO-I models to estimate their SLO fulfillment according to the combined hardware load (Lines 9 \& 10). The last step is calculating the offloading gain ($\gamma$) for each platoon member ($w$), i.e., whether global SLO fulfillment would be improved by offloading $s$ to $w$, and then return the best possible vehicle. For this, it first calculates $\gamma$ (Line 11), which is appended to the collection $\Gamma$. In the final step, it selected the best alternative host among the platoon members (Line 14); however, if not even the best host would improve overall SLO fulfillment, it prefers to keep the current host (Line 15). The outcome is returned to Algo.~\ref{alg:observation}, which offloads the service accordingly.






\section{Evaluation}\label{sec:evaluation}
In the following section, we describe how the presented methodology was implemented and evaluated for a set of heterogeneous perception services and a composable vehicle platoon. For this, we implement a prototype of our framework that addresses the illustrated scenario; afterward, we document the experimental setup, including service implementations and applied processing hardware, then present the experimental results, and conclude with a critical discussion.

\subsection{Implementation}
\label{subsec:implementation}


To implement our methodology, we provide a Python-based prototype\footnotemark that follows a clear modular structure for services, their SLOs, and device types.
\footnotetext{The framework prototype is available at \href{https://anonymous.4open.science/r/intelligentVehicle-720C/}{GitHub}, accessed on July 14th 2024}
Hence, the framework can be extended with new services as long as they are supported by the underlying edge device. Once the framework is installed\footnotemark, services can be started or stopped remotely through HTTP; for running the experiments, we send the respective instructions to different platoon members using Postman flows\footnote{Postman is a common tool for sending HTTP requests; Postman flows is a UI extension that allows to specify sequences of requests, e.g., start/stop services}.
To isolate resource consumption, services are executed in individual Python threads. During that time, each service observes its own SLO fulfillment as part of its service wrapper (i.e., Algo.~\ref{alg:observation}); in the present state, this is done every 500ms, though it can be customized for service types or instances.
To avoid interfering with regular service execution, model training and evaluation of alternative service hosts run detached from the main service thread.
\footnotetext{Instructions are provided in the following \href{https://anonymous.4open.science/r/intelligentVehicle-720C/README.md}{README}, accessed on July 14th 2024}

Vehicles communicate exclusively over HTTP; the respective connection is established either through a local access point managed by the platoon leader, or through IBSS, i.e., a peer-to-peer network. 
Training and updating of SLO-I models, or rather their underlying BNs, uses pgmpy~\cite{ankan_pgmpy_2023}, a Python library for Bayesian Network Learning (BNL).
In pgmpy, BNs can be encoded in XML, which each had a size of roughly 10kB in our evaluation; hence, a feasible size to be transmitted and shared within the platoon.

\subsection{Experimental Setup}
\label{subsec:experimental-setup}

To evaluate our prototype in a realistic environment, we implement the scenario illustrated in Section~\ref{subsec:illustrative-scenario}, i.e., perception services are offloaded within a vehicle platoon according to their local SLO fulfillment. We provide three perception services that can be executed on edge devices; Tab.~\ref{tab:service-list} provides essential information on these services: \textit{CV} uses Yolov8 to detect objects in a video stream, \textit{LI} processes point clouds from a Lidar sensor to map the environment, and \textit{QR} uses OpenCV to detects QR codes in a video. Each service has specific tuning parameters, such as the resolution (\textit{pixel}) and \textit{fps} for \textit{CV} and \textit{QR}; \textit{LI} accepts an additional parameter \textit{mode} to define the point cloud radius. 

According to our expert knowledge, each service's expected QoS level is specified through a list of SLOs; through heuristic trial and error, the following ones proved especially useful: we constrain the processing $\textbf{time} \leq 1000 / \textit{fps}$, i.e., frames must be processed faster than they come in; the maximum \textbf{energy} consumption can be adjusted for individual devices: we put a limit of $\leq 15 W$ for regular platoon members and $\leq 25 W$ for the platoon leader. Notice, that this considers the vehicle-wide energy consumption over all executed services. According to the video resolution (\textit{pixel}) provided to \textit{CV}, the service uses the respective Yolov8 model size (i.e., v8n, v8s, v8m); however, this affects the number of objects that are detected, which is ensured through the \textbf{rate} SLO. 


\setlength{\tabcolsep}{5pt}
\begin{table}[t]
    \small
  \centering
  \caption{List of all predefined services that were added to the framework}
  \label{tab:service-list}
  \begin{tabular}{clccc}
    \toprule
    ID & Service Description & CUDA & Parameters & SLOs\\
    \midrule

    \textit{CV} & Object Detection with Yolov8~\cite{varghese_yolov8_2024} & Yes & \textit{pixel}, \textit{fps} &  \textbf{time}, \textbf{energy}, \textbf{rate}\\
    \textit{LI} & Lidar Point Cloud Processing~\cite{dzung_maudzungsfa3d_2020} & Yes & \textit{mode}, \textit{fps} & \textbf{time}, \textbf{energy} \\
    \textit{QR} & Detect QR Code w/ OpenCV~\cite{opencv_opencv_2024} & No & \textit{pixel}, \textit{fps} & \textbf{time}, \textbf{energy} \\
    \bottomrule
  \end{tabular}
  \vspace{-10pt}
\end{table}

\begin{figure}[t]
\centering
\subfloat[CV (Yolov8)]{\label{subfig:demo-cv}\centering\includegraphics[width=.325\columnwidth]{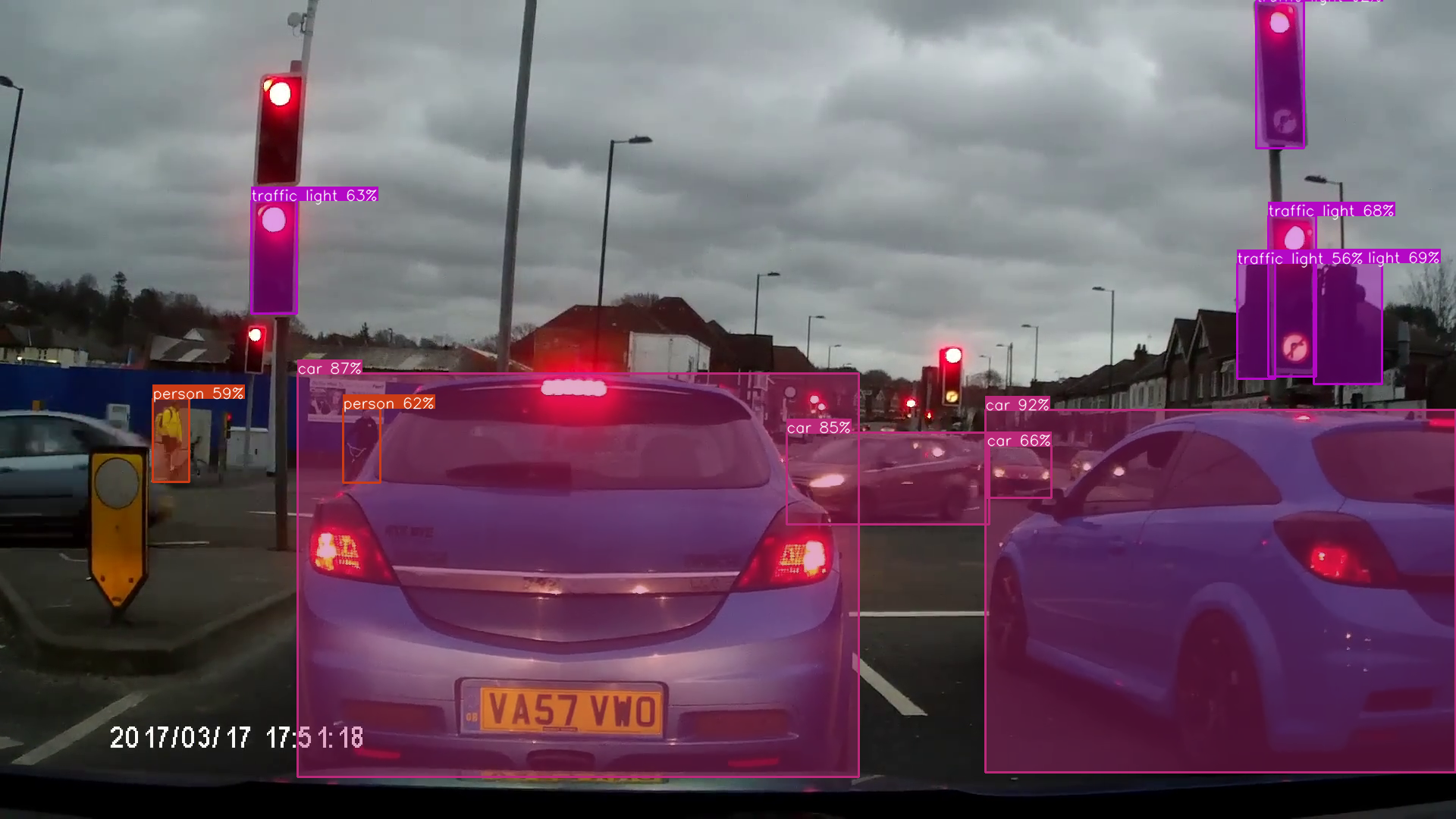}}
\hspace{0.005\columnwidth}
\subfloat[LIdar (SFA3D)]{\label{subfig:demo-li}\centering\includegraphics[width=.324\columnwidth]{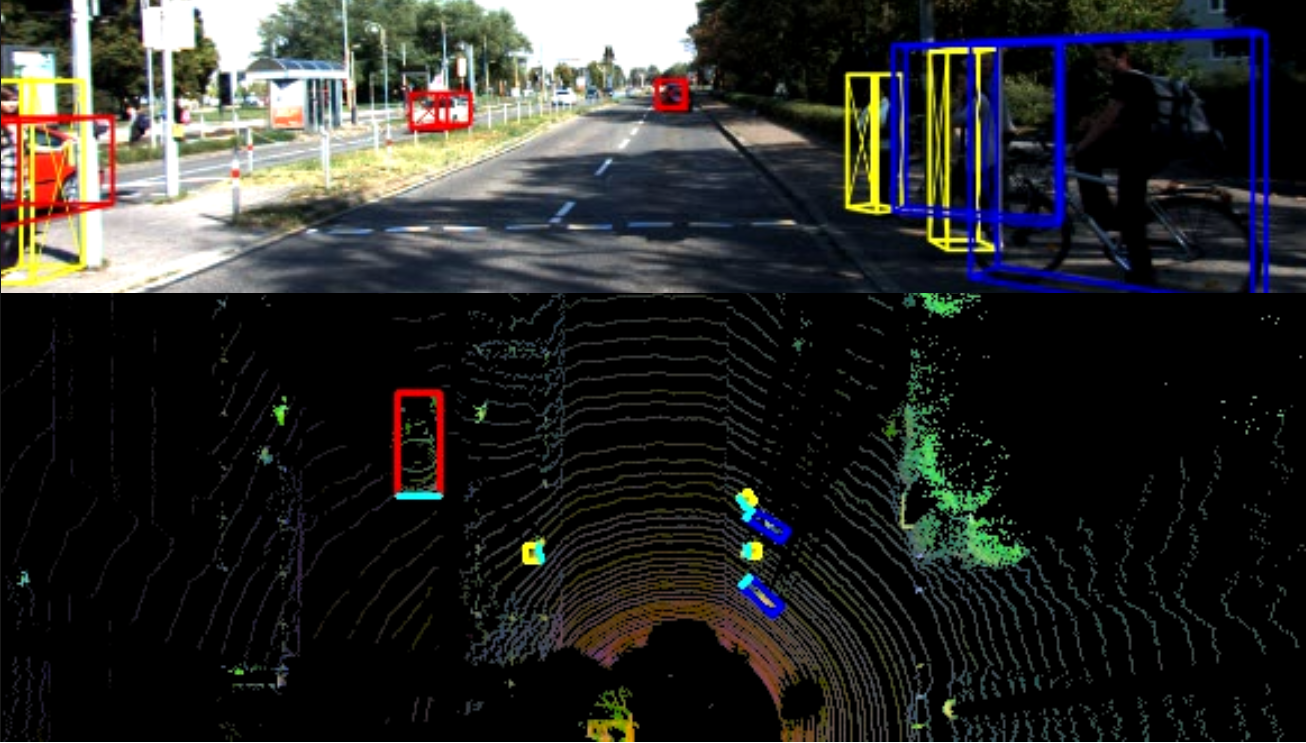}}
\hspace{0.005\columnwidth}
\subfloat[QR (OpenCV)]{\label{subfig:demo-li}\centering\includegraphics[width=.314\columnwidth]{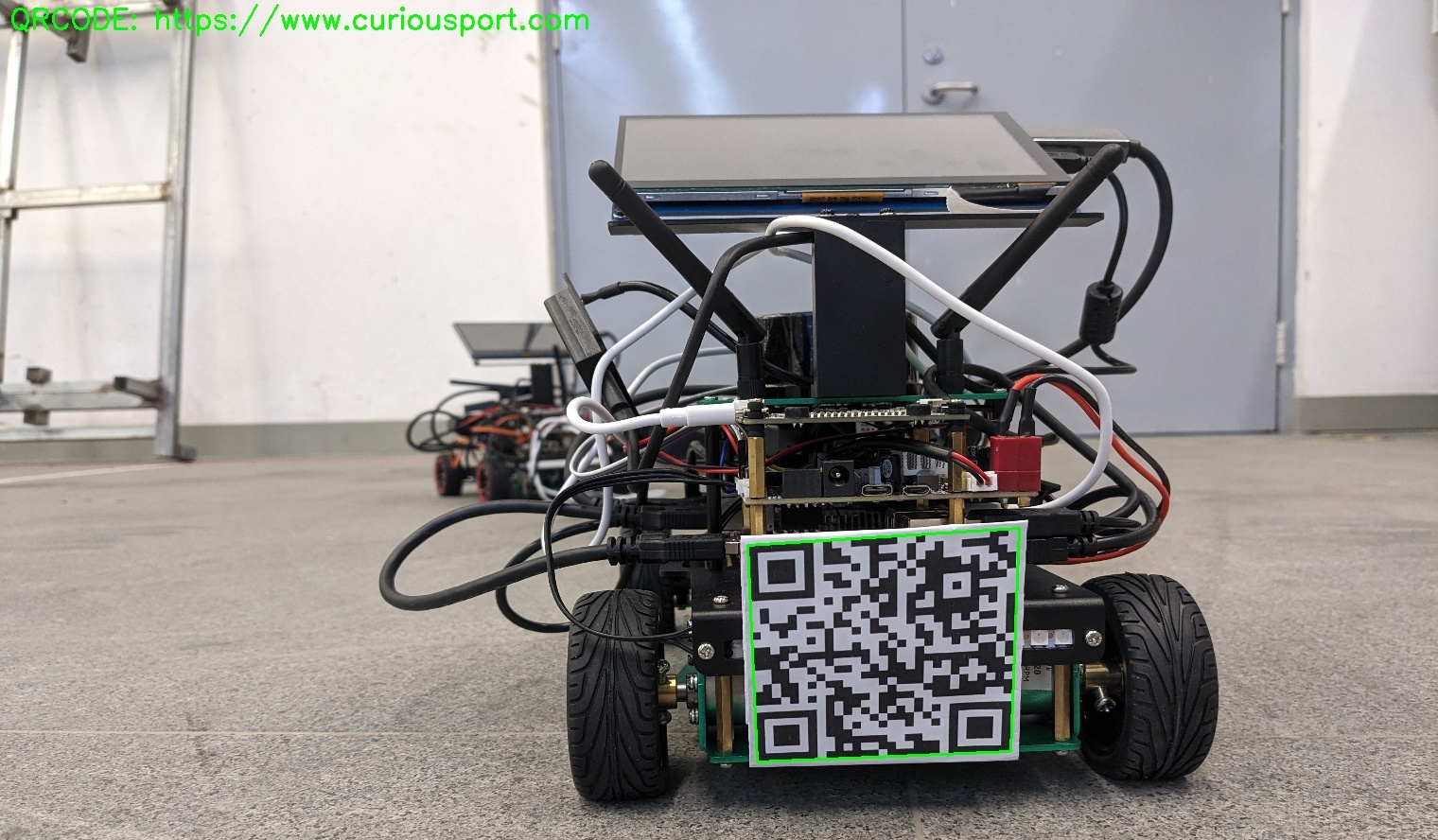}}
\caption{Demo output for each service according to the prerecorded input data}
\label{fig:services-demo}
\vspace{-15pt}
\end{figure}

The presented framework is evaluated on two different instances of Nvidia Jetson boards, namely Jetson Orin \textit{NX} and Orin \textit{AGX}, which are described in more detail in Tab.~\ref{tab:device-list}: the \textit{AGX} is clearly superior in terms of memory and GPU and has a slightly better CPU. While the specific Nvidia CUDA version has minor importance, CUDA itself is crucial to accelerate the \textit{CV} and \textit{LI} services. Each Jetson \textit{NX} is embedded in a Rosmaster R2\footnotemark car -- a battery-powered multi-sensory vehicle used for development. To ensure a stable evaluation environment, the service processed either prerecorded videos (\textit{CV} \& \textit{QR}) or binary-encoded point clouds ($LI$); Fig.~\ref{fig:services-demo} shows a demo output for each service.
\footnotetext{More information about the Rosmaster R2 \href{https://github.com/YahboomTechnology/ROSMASTER-R2}{here}, accessed Jul 14th 2024}

\setlength{\tabcolsep}{4.5pt}
\begin{table}[t]
    \small
  \centering
  \caption{List of all edge devices that were involved in the evaluation}
  \label{tab:device-list}
  \begin{tabular}{lcrcrcr}
    \toprule
    Full Device Name & ID & Price\footnotemark & CPU & RAM &  GPU & CUDA \\
    \midrule

    Jetson Orin NX ($3$)  & \textit{NX} & 450 €  & ARM Cortex 8C  & 8 GB  & Volta 1k & 11.4\\
    Jetson Orin AGX   & \textit{AGX} & 800 €  & ARM Cortex 12C  & 64 GB  & Volta 2k & 12.2\\
    \bottomrule
  \end{tabular}
  \vspace{-10pt}
\end{table}
\footnotetext{Prices adopted from \href{https://sparkfun.com/}{sparkfun}, accessed Jul 14th 2024}

\subsection{Results}
\label{subsec:results}

Given the experimental setup, we evaluate the prototype based on: (1) what is the overhead of continuously interpreting services, and what limitations arise from the platoon size; (2) can SLO-aware retraining ensure prediction accuracy regardless of unexpected runtime behavior; and (3) can the framework fulfill high-level SLOs within the platoon by offloading computations?
In the following, we evaluate these aspects using two base cases and one advanced scenario. For each case, we describe the vehicle operations and the experimental results:

\vspace{-10pt}
\subsubsection{Scenario 1A} An individual vehicle (i.e., \textit{NX} or \textit{AGX}) executes the \textit{QR} service; every 25s, we add a vehicle to its platoon, up to a maximum size of 4 vehicles. Given this, we track the time to execute the service wrapper, i.e., how long it takes to retrain the SLO-I model and evaluate alternative hosts for \textit{QR}.

\begin{figure}[t]
\centering
\subfloat[Orin \textit{NX}]{\label{subfig:overhead-nx}\centering\includegraphics[width=.5\columnwidth]{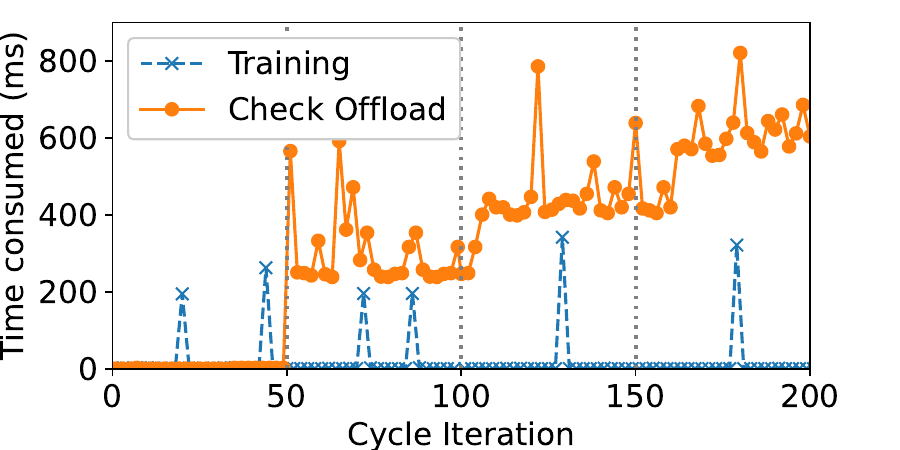}}
\subfloat[Orin \textit{AGX}]{\label{subfig:overhead-agx}\centering\includegraphics[width=.5\columnwidth]{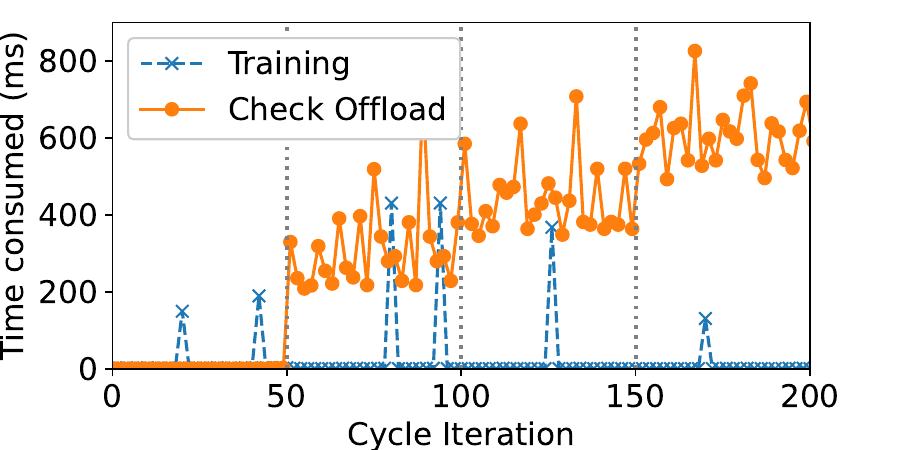}}
\hfill
\vfill
\caption{Time required to train the SLO-I model and evaluate alternative hosts}
\label{fig:wrapper-overhead}
\vspace{-11pt}
\end{figure}

Fig.~\ref{fig:wrapper-overhead} visualizes the times required to train the SLO-I model or evaluate alternative hosts for offloading; both processes are executed as part of the service wrapper. The wrapper runs every 500ms for a total of 100 seconds, hence, the plot contains 200 wrapper iterations. Vertical grey lines indicate when an additional device is introduced to the platoon, i.e., at 50, 100, and 150 iterations.

Given this, we conclude that the platoon size has a linear impact on the time required to evaluate alternative hosts; the exception is $|P| = 1$, when evaluating other vehicles for offloading is obsolete. For a platoon with $|P| \leq 3$, the entire service wrapper finished mostly in $\leq 500 ms$; however, $|P| \geq 4$ starts exceeding $500ms$, which indicates that it would not be possible to interpret the SLO fulfillment every $500ms$. This could be overcome by either structuring the platoon into smaller subgroups or adjusting the evaluation interval.


\vspace{-11pt}
\subsubsection{Scenario 1B} An individual vehicle (i.e., AGX) runs \textit{CV} locally; however, the respective SLO-I model was not yet fine-tuned and initial predictions are likely inaccurate. Additionally, variable drifts occur, which we simulate through stress-ng: after $125s$ the CPU load of \textit{AGX} is stressed 40\%. We measure $p_\phi$ and $W_\phi$, and compare our presented training strategy with a static service wrapper.

\begin{figure}[t]
\centering
\subfloat[Without SLO-dependent retraining]{\label{subfig:overhead-nx}\centering\includegraphics[width=.5\columnwidth]{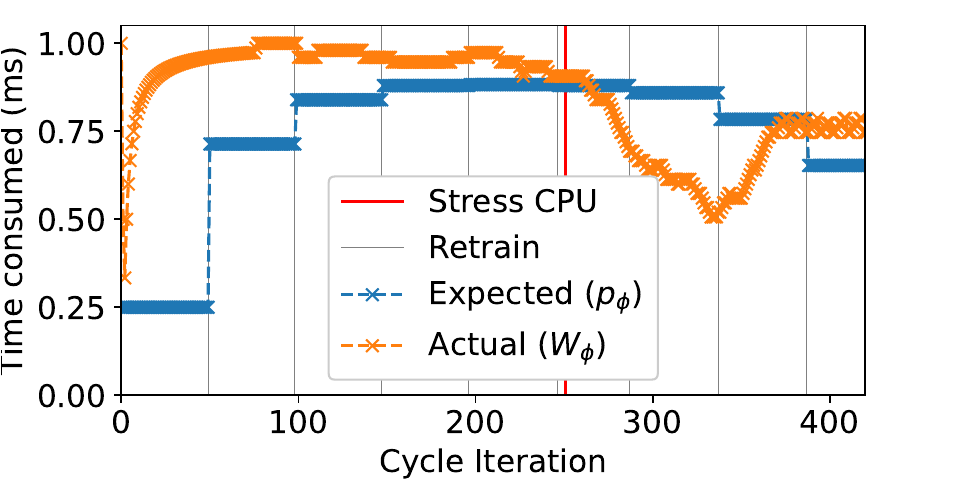}}
\subfloat[With SLO-dependent retraining]{\label{subfig:overhead-agx}\centering\includegraphics[width=.5\columnwidth]{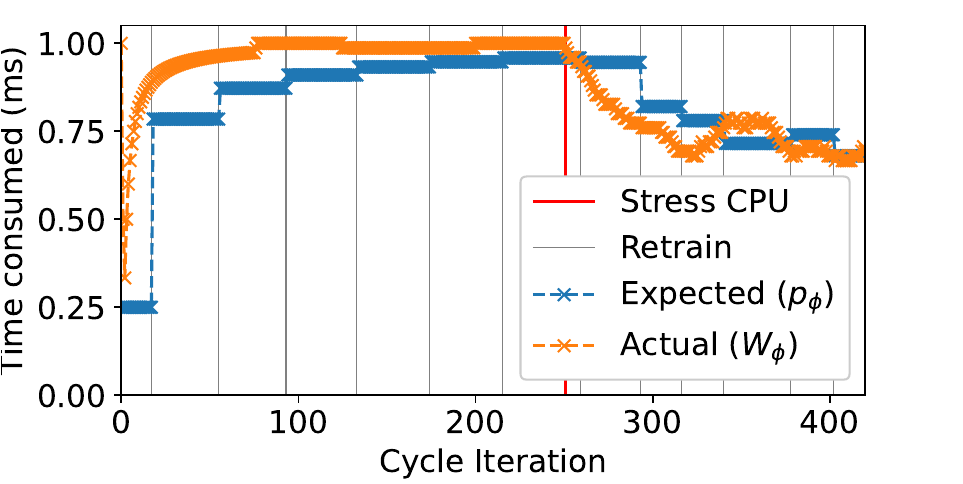}}
\hfill
\vfill
\caption{Improved prediction accuracy through SLO-dependent retraining}
\label{fig:smooth-retraining}
\end{figure}

Fig.~\ref{fig:smooth-retraining} visualizes for both runs the predicted ($p_\phi$) and actual SLO fulfillment ($W_\phi$); vertical grey lines indicate when retraining happened, and the red line when the perturbation occurred. Not only does the left side perform fewer retraining, i.e., 8 instead of 12, but more importantly, the right side presents shorter training intervals when the SLO fulfillment is unstable, such as during the period between $x=[250,350]$. Consequentially, the Mean Squared Error (MSE) was $0.07$ on the left and $0.01$ on the right side; given that, we conclude that SLO-dependent retaining helped to increase the prediction accuracy for initially inaccurate models or at runtime when perturbations occur.

\begin{figure}[t]
\centering
\includegraphics[width=0.98\columnwidth]{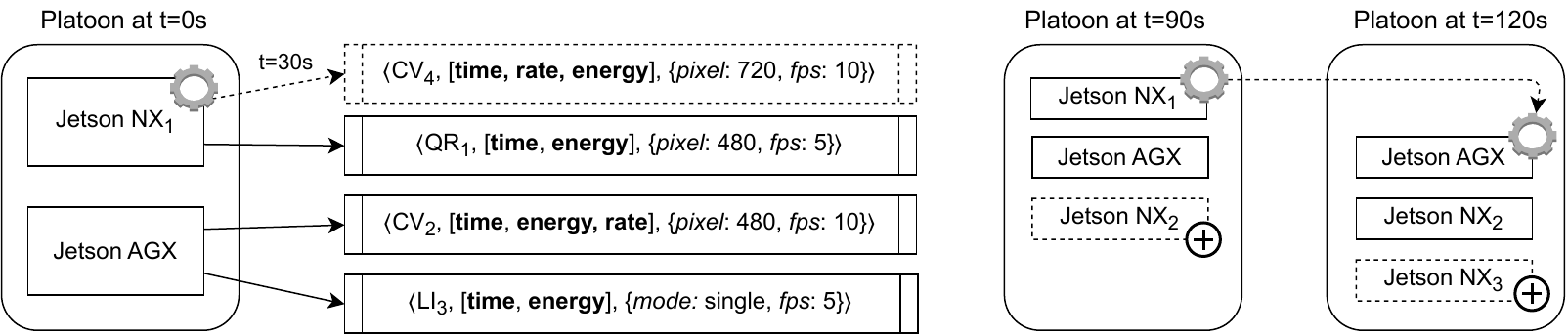}
\caption{Sequential description of Scenario 2: starting services and adjusting the platoon}
\label{fig:scenario-desc}
\vspace{-10pt}
\end{figure}

\vspace{-11pt}
\subsubsection{Scenario 2} Fig.~\ref{fig:scenario-desc} provides a sequential description of this scenario: at time $t=0s$ the platoon $P = \{NX_1, AGX\}$ starts 3 services (i.e., $QR_1$, $CV_2$, $LI_3$); at $t = 30s$ \textit{$NX_1$} starts $CV_4$; at $t=90s$ \textit{$NX_2$} joins the platoon, and at $t=120s$ \textit{$NX_3$} joins, \textit{$NX_1$} leaves the platoon, and leadership is transferred to \textit{AGX}.

\begin{figure}[t]
\vspace{-6pt}
\centering
\vfill
\subfloat[Orin \textit{$NX_1$}]{\label{subfig:slo-f-nx1}\centering\includegraphics[width=.385\columnwidth]{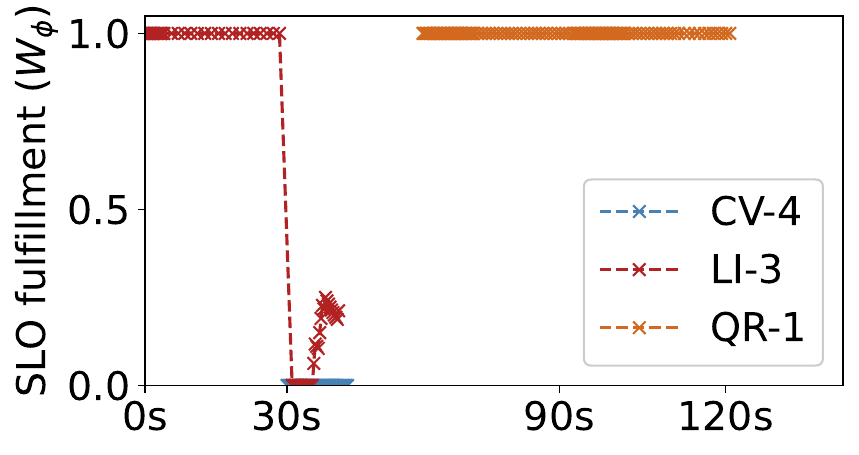}}
\subfloat[Orin \textit{$AGX$}]{\label{subfig:slo-f-agx}\centering\includegraphics[width=.385\columnwidth]{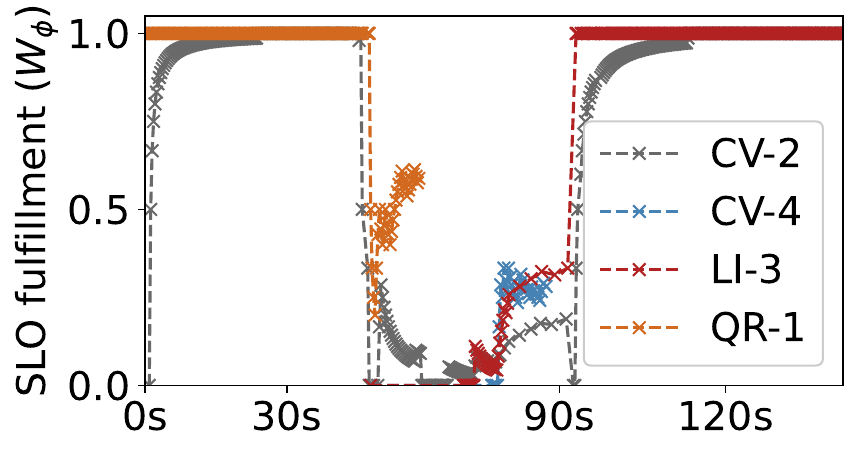}}
\hspace{0.01\columnwidth}
\subfloat[$CV_4$ at $t=90s$]{\label{subfig:slo-f-t90}\centering\includegraphics[width=.21\columnwidth]{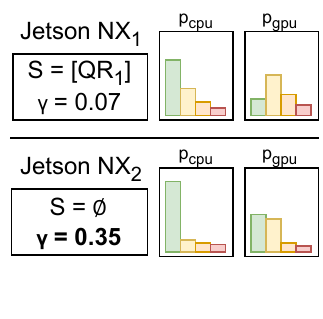}}
\caption{SLO fulfillment and decision making for constrained services in the platoon}
\label{fig:slo-f}
\end{figure}

Fig.~\ref{fig:slo-f} visualizes the SLO fulfillment of all services executed at \textit{$NX_1$} and \textit{AGX}; at first, all three services (i.e., $QR_1$, $CV_2$, $LI_3$) achieve maximum SLO fulfillment, i.e., $W_\phi = 1.0$. However, as soon as $CV_4$ is started at $t=30s$, $NX_1$ fails to ensure the SLOs for both $LI_3$ and $CV_4$. Due to that, $NX_1$ decides to load off both services to \textit{AGX}, which in turn, causes \textit{AGX} to fail most of its services' SLOs. 
This changes at $t=55s$, when \textit{AGX} decides to move one of its services (i.e., \textit{QR-1}) to \textit{$NX_1$}, which slightly recovers the SLO fulfillment of the remaining three services. Next, at $t=90s$, $NX_2$ joins the platoon, which encourages \textit{AGX} to offload another service (i.e., $CV_4$) to $NX_2$. Here, Fig.~\ref{subfig:slo-f-t90} shows the decision-making of \textit{AGX}: since $NX_1$ already executes $QR_1$, it estimates how adding $CV_4$ would have a negative impact on $QR_1$ due to predicted resource shortage; hence, it chooses $NX_2$, which promises global SLO improvement of $\gamma = 0.35$.

Given this, we conclude that services can react in $\leq 10s$ to local SLO violations, which appears practical for real-time systems. This highlights the impact of co-locating too many services at one edge device and how this can be resolved by adding new vehicles to the platoon. Furthermore, changing the platoon leader at $t=120$ showed no negative impact on the remaining vehicles -- its ongoing computations were shifted to an idle vehicle (i.e., $NX_4$) that just had joined.

\section{Conclusion \& Future Work}\label{sec:conclusion}
This paper introduced a novel V2V offloading mechanism that ensures high-level requirements during runtime. By leveraging probabilistic models, our approach predicts how offloading a service would impact platoon-wide SLO fulfillment. Thus, individual services have the capability to choose an alternative host, if this promises to improve SLO fulfillment.  
Notably, our evaluation demonstrated how the proposed framework could handle both an increasing number of platoon members as well as a series of heterogeneous services. In practice, thanks to our real-case test bed, our analysis highlights how our approach can improve real-time decision-making within an AV platoon, making it a valuable extension.
While our work has shown promising results, there are still limitations and areas of improvement: first, our work does not consider network latency; while we ruled it negligible in our case, future work can include more detailed analyses. Furthermore, our implementation executes services in Python threads; we plan to implement a more effective and elegant solution, containerizing each service instance. Another interesting direction would be to explore more complex scenarios in which a single platoon has multiple swarms or when multiple platoons need to coordinate with each other. In this case, it might be necessary and helpful to perform a hyperparameter optimization for the presented methodology.

\bibliographystyle{splncs04}
\bibliography{boris,andrea}

\begin{thebibliography}{10}
\providecommand{\url}[1]{\texttt{#1}}
\providecommand{\urlprefix}{URL }
\providecommand{\doi}[1]{https://doi.org/#1}

\bibitem{ankan_pgmpy_2023}
Ankan, A., Textor, J.: pgmpy: {A} {Python} {Toolkit} for {Bayesian} {Networks} (Apr 2023)

\bibitem{bacchus_representing_1989}
Bacchus, F.I.: Representing and reasoning with probabilistic knowledge. Artificial {Intelligence}, MIT Press (1989)

\bibitem{chen2020delay}
Chen, C., Chen, L., Liu, L., He, S., Yuan, X., Lan, D., Chen, Z.: Delay-optimized v2v-based computation offloading in urban vehicular edge computing and networks. IEEE Access  \textbf{8},  18863--18873 (2020)

\bibitem{dong_quantum_2023}
Dong, S., Xia, Y., Kamruzzaman, J.: Quantum {Particle} {Swarm} {Optimization} for {Task} {Offloading} in {Mobile} {Edge} {Computing}. IEEE Transactions on Industrial Informatics  \textbf{19}(8),  9113--9122 (Aug 2023). \doi{10.1109/TII.2022.3225313}

\bibitem{du_cooperative_2020}
Du, H., Leng, S., Zhang, K., Zhou, L.: Cooperative {Sensing} and {Task} {Offloading} for {Autonomous} {Platoons}. In: {IEEE} {GLOBECOM} 2020 (Dec 2020)

\bibitem{dzung_maudzungsfa3d_2020}
Dzung, N.M.: maudzung/{SFA3D} (2020), \url{https://github.com/maudzung/SFA3D}

\bibitem{fan2023joint}
Fan, W., Su, Y., Liu, J., Li, S., Huang, W., Wu, F., Liu, Y.: Joint task offloading and resource allocation for vehicular edge computing based on v2i and v2v modes. IEEE Transactions on Intelligent Transportation Systems  (2023)

\bibitem{fan_minimum-cost_2019}
Fan, X., Cui, T., Cao, C., Chen, Q., Kwak, K.S.: Minimum-{Cost} {Offloading} for {Collaborative} {Task} {Execution} of {MEC}-{Assisted} {Platooning}. Sensors  (2019)

\bibitem{gao2021federated}
Gao, Y., Liu, L., Zheng, X., Zhang, C., Ma, H.: Federated sensing: Edge-cloud elastic collaborative learning for intelligent sensing. IEEE Internet of Things Journal  \textbf{8}(14),  11100--11111 (2021)

\bibitem{guo_toward_2020}
Guo, H., Liu, J., Lv, J.: Toward {Intelligent} {Task} {Offloading} at the {Edge}. IEEE Network  \textbf{34}(2),  128--134 (Mar 2020). \doi{10.1109/MNET.001.1900200}

\bibitem{guo2022v2v}
Guo, H., Rui, L.l., Gao, Z.p.: V2v task offloading algorithm with lstm-based spatiotemporal trajectory prediction model in svcns. IEEE Transactions on Vehicular Technology  \textbf{71}(10),  11017--11032 (2022)

\bibitem{hazra_cooperative_2023}
Hazra, A., Donta, P.K., Amgoth, T., Dustdar, S.: Cooperative {Transmission} {Scheduling} and {Computation} {Offloading} {With} {Collaboration} of {Fog} and {Cloud} for {Industrial} {IoT} {Applications}. IEEE Internet of Things Journal  (Mar 2023)

\bibitem{hu_task_2019}
Hu, Y., Cui, T., Huang, X., Chen, Q.: Task {Offloading} {Based} on {Lyapunov} {Optimization} for {MEC}-assisted {Platooning}. In: 2019 11th {International} {Conference} on {Wireless} {Communications} and {Signal} {Processing} ({WCSP}). pp.~1--5 (Oct 2019)

\bibitem{kuutti2020survey}
Kuutti, S., Bowden, R., Jin, Y., Barber, P., Fallah, S.: A survey of deep learning applications to autonomous vehicle control. IEEE Transactions on Intelligent Transportation Systems  \textbf{22}(2),  712--733 (2020)

\bibitem{le2022survey}
Le~Mero, L., Yi, D., Dianati, M., Mouzakitis, A.: A survey on imitation learning techniques for end-to-end autonomous vehicles. IEEE Transactions on Intelligent Transportation Systems  \textbf{23}(9),  14128--14147 (2022)

\bibitem{lu_cooperative_2022}
Lu, L., Li, X., Sun, J., Yang, Z.: Cooperative {Computation} {Offloading} and {Resource} {Management} for {Vehicle} {Platoon}: {A} {Deep} {Reinforcement} {Learning} {Approach}. In: {IEEE} {Int} {Conf} on {High} {Performance} {Computing} \& {Communications} (2022)

\bibitem{ma_edge_2022}
Ma, X., Su, Z., Xu, Q., Ying, B.: Edge {Computing} and {UAV} {Swarm} {Cooperative} {Task} {Offloading} in {Vehicular} {Networks}. In: 2022 {International} {Wireless} {Communications} and {Mobile} {Computing} ({IWCMC}). pp. 955--960 (May 2022)

\bibitem{ma2020artificial}
Ma, Y., Wang, Z., Yang, H., Yang, L.: Artificial intelligence applications in the development of autonomous vehicles. IEEE Journal of Automatica Sinica  (2020)

\bibitem{mousa_efficient_2022}
Mousa, M.H., Hussein, M.K.: Efficient {UAV}-based mobile edge computing using differential evolution and ant colony optimization. PeerJ Computer Science  (2022)

\bibitem{opencv_opencv_2024}
{opencv}: opencv at 4.9.0 (2024), \url{https://github.com/opencv/opencv/tree/4.9.0}

\bibitem{qiao2018collaborative}
Qiao, G., Leng, S., Zhang, K., He, Y.: Collaborative task offloading in vehicular edge multi-access networks. IEEE Communications Magazine  \textbf{56}(8),  48--54 (2018)

\bibitem{sedlak_equilibrium_2024}
Sedlak, B., Pujol, V.C., Donta, P.K., Dustdar, S.: Equilibrium in the {Computing} {Continuum} through {Active} {Inference}. Future Generation Computer System  (2024)

\bibitem{sedlak_diffusing_2024}
Sedlak, B., Pujol, V.C., Donta, P.K., Dustdar, S.: Diffusing {High}-level {SLO} in {Microservice} {Pipelines}. In: 2024 {IEEE} {International} {Conference} on {Service}-{Oriented} {System} {Engineering} ({SOSE}) (2024)

\bibitem{tasoulas_bayllocator_2012}
Tasoulas, V., Haugerud, H., Begnum, K.M.: Bayllocator: {A} {Proactive} {System} to {Predict} {Server} {Utilization} and {Dynamically} {Allocate} {Memory} {Resources} {Using} {Bayesian} {Networks} and {Ballooning} (Dec 2012)

\bibitem{varghese_yolov8_2024}
Varghese, R., M., S.: {YOLOv8}: {A} {Novel} {Object} {Detection} {Algorithm} with {Enhanced} {Performance} and {Robustness}. In: {ADICS} (2024)

\bibitem{wu_intelligent_2023}
Wu, Y., Cai, C., Bi, X., Xia, J., Gao, C., Tang, Y., Lai, S.: Intelligent resource allocation scheme for cloud-edge-end framework aided multi-source data stream. EURASIP Journal on Advances in Signal Processing  \textbf{2023} (May 2023)

\bibitem{xu_optimization_2023}
Xu, B., Deng, T., Liu, Y., Zhao, Y., Xu, Z., Qi, J., Wang, S., Liu, D.: Optimization of cooperative offloading model with cost consideration in mobile edge computing. Soft Computing  \textbf{27}(12),  8233--8243 (Jun 2023)

\bibitem{zhang_distributed_2023}
Zhang, Z., Jiang, J., Xu, H., Zhang, W.A.: Distributed dynamic task allocation for unmanned aerial vehicle swarm systems: {A} networked evolutionary game-theoretic approach. Chinese Journal of Aeronautics  (Dec 2023)

\end{thebibliography}
\end{document}